# Automated Diabetic Retinopathy Grading using Deep Convolutional Neural Network

Saket S. Chaturvedi*, Kajol Gupta, Vaishali Ninawe, Prakash S. Prasad*

Department of Computer Science & Engineering, Priyadarshini Institute of Engineering & Technology, Nagpur, India

Email: saketschaturvedi@gmail.com, prakashsprasad@gmail.com

**ABSTRACT**

Diabetic Retinopathy is a global health problem, influences 100 million individuals worldwide, and in the next few decades, these incidences are expected to reach epidemic proportions. Diabetic Retinopathy is a subtle eye disease that can cause sudden, irreversible vision loss. The early-stage Diabetic Retinopathy diagnosis can be challenging for human experts, considering the visual complexity of fundus photography retinal images. However, Early Stage detection of Diabetic Retinopathy can significantly alter the severe vision loss problem. The competence of computer-aided detection systems to accurately detect the Diabetic Retinopathy had popularized them among researchers. In this study, we have utilized a pre-trained DenseNet121 network with several modifications and trained on APTOS 2019 dataset. The proposed method outperformed other state-of-the-art networks in early-stage detection and achieved 96.51% accuracy in severity grading of Diabetic Retinopathy for multi-label classification and achieved 94.44% accuracy for single-class classification method. Moreover, the precision, recall, f1-score, and quadratic weighted kappa for our network was reported as 86%, 87%, 86%, and 91.96%, respectively. Our proposed architecture is simultaneously very simple, accurate, and efficient concerning computational time and space.

## 1. INTRODUCTION

The incidence of vision loss due to Diabetic Retinopathy is on the rise, and in the next few decades, these incidences are expected to reach epidemic proportions globally. Since 1980, the cases of diabetes prevalence have quadrupled [1], [2]. The visual impairment and blindness estimates for Diabetic Retinopathy increased by 64% and 27% between 1990 and 2010, respectively [3]. In 2017, 425 million people worldwide were reported with diabetes, and this number is estimated to increase to 642 million by 2040 [4]. It is estimated that nearly every patient with type-1 diabetes and 60% of patients with type-2 diabetes would develop Diabetic Retinopathy in the first 20 years of diabetes [5], [6].

Diabetic Retinopathy (DR) is the most mundane and subtle microvascular complication of diabetes, resulting in a sudden loss of vision. Diabetic Retinopathy often remains undetected until it progresses to an advanced vision-threatening stage as this complicated problem can only be noticed when the tiny blood vessels in the retina begin to damage. This tiny blood vessel causes the blood flow, and fluid present in the retina results in forming features. After the disease starts growing to the next level, oxygen enters in between the retina and clouding vision because of the generation of new blood vessels. In the case of diabetic patients, it is essential to conduct regular screening to track the growth of DR [7] among four severity levels: Normal, Mild, Moderate, Severe, and Proliferative

Diabetic Retinopathy. The most dangerous stage of Diabetic Retinopathy is the proliferative DR in which the likelihood of blood leaking is at a peak, causing permanent vision loss.

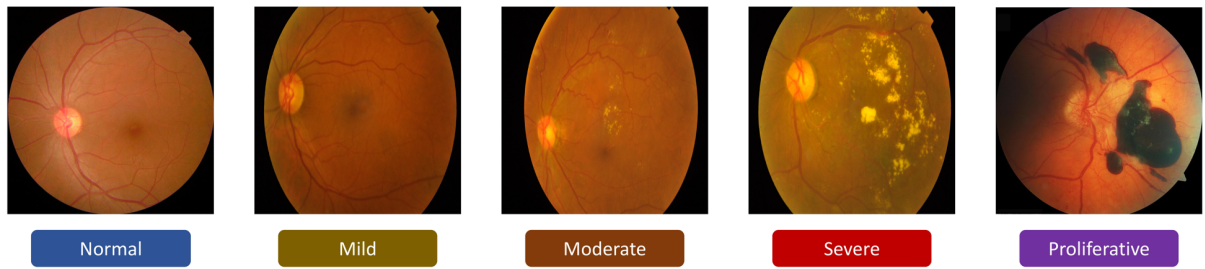

| Normal | Mild | Moderate | Severe | Proliferative |

**Figure 1.** The fundus photography images from APTOS2019 Kaggle's dataset at each of the Normal, Mild, Moderate, Severe, and Proliferative Diabetic Retinopathy severity levels.

The current state of Diabetic Retinopathy screening in the real world is based on the assessment of color fundus photography (see Figure 1), which is induced by an Ophthalmologist. The fundus photography leaves a large proportion of patients undiagnosed and therefore receiving medical help too late, owing to low adherence and access to retina screening visits [8]. In-person expert examinations are impractical and unsustainable, given the pandemic size of the diabetic population [9]–[11]. However, it is time-consuming and resource-demanding to grade the images manually. Notwithstanding, early detection and prevention of DR progression are essential to mitigate the rising threat of DR.

The presence of an automated or computer-aided system can make it very easy for a specialist to observe the retina of diabetic patients clearly [12]. Artificial intelligence (AI) may offer a solution to this conundrum. Deep Learning (DL), and correctly, Deep Convolutional Neural Networks (DCNNs) [13], can be used for an end-to-end assessment of raw medical images to produce a target outcome prediction. The diagnostic use of DCCN algorithms is already spreading in various healthcare areas [14],[15], such as radiology, dermatology [16], and pathology [17]. In ophthalmology, ground-breaking work has recently been conducted on the automation of DR grading [18]–[20].

## 2. RELATED WORKS

The computer-aided detection systems capability to accurately detect the grades of Diabetic Retinopathy had made them popular among researchers. In the last ten years, numerous research work focusing on the development of Computer-Aided systems to automatically detect Diabetic Retinopathy using traditional machine learning algorithms were recorded.

Quellec et al. [21] used a traditional KNN algorithm with optimal filters on two classes to achieve an AUC of 0.927. Also, Sinthanayothin et al. [22] proposed an automated Diabetic Retinopathy detection system on morphological features using the KNN algorithm and obtained sensitivity and specificity of 80.21% and 70.66%, respectively. Further, In the paper [23], three classes of Diabetic Retinopathy were classified using Neural Network. They classified mild, moderate, and severe stages of Diabetic Retinopathy with an accuracy of 82.6%, 82.6%, and 88.3%, respectively.

Larsen et al. [24] demonstrated an automatic diagnosis of Diabetic Retinopathy in fundus photographs with a visibility threshold. They reported an accuracy of 90.1% for true cases detection and 81.3% for the detection of the false case. Agurto et al. [25] utilized multi-scale Amplitude Modulation and Frequency Modulation based decomposition to distinguish between Diabetic

Retinopathy and normal retina images. In [26], the authors reported an area under ROC of 0.98 for Texture features and accuracy of 99.17% for two-class classification by using Wavelet transform with SVM.

Jelinek et al. [27] proposed an automated Diabetic Retinopathy detection by combining the works of Spencer [28] and Cree [27] system, which achieved a sensitivity of 85% and specificity of 90%. Abràmo et al. [29] developed the Eye-Check algorithm for automated Diabetic Retinopathy detection. They detected abnormal lesions with an AUC of 0.839. Dupas et al. [30] developed a Computer-Aided Detection system with a KNN classifier to detect Diabetic Retinopathy with a sensitivity of 83.9% and specificity of 72.7%.

Acharya et al. [31] classified five classes using SVM classifier on the bi-spectral invariant features to achieve sensitivity, specificity, and accuracy of 82%, 86%, and 85.9%, respectively. They also worked utilizing four features and achieved a classification accuracy of 85%, sensitivity of 82%, and specificity of 86%. Roychowdhury et al. [32] proposed a two-step classification approach. In the first step, the false positives were removed. Later, GMM, KNN, and SVM were utilized for the classification task. They reported a sensitivity of 100%, a specificity of 53.16%, and AUC of 0.904.

Deep learning algorithms have become popular in the last few years. Kaggle [33] has launched several competitions focusing on automated grading of Diabetic Retinopathy detection. Pratt et al. [34] introduced a CNN based method, which even surpassed human experts in the classification of advanced stage Diabetic Retinopathy. Kori et al. [35] utilized an ensemble of ResNet and Densely connected networks to detect advanced stages of Diabetic Retinopathy and macular enema. Torrey et al. [36] developed a more interpretable CNN model to detect a lesion in the retinal fundus images. In a similar study [37], further advancement in the automated Diabetic Retinopathy methods was done along with RAM. Yang et al. [38] employed an unbalanced weight map methodology to emphasize lesion detection with an AUC of 0.95. In [39], VGG-16 and Inception- 4 networks were utilized for effective Diabetic Retinopathy classification.

Previous studies had mostly focused on the binary classification of Diabetic Retinopathy, which restricted the scope of Diabetic Retinopathy Detection studies. The purpose of this work was to predict the severity level of Diabetic Retinopathy fundus photography images among five classes–No DR, Mild DR, Moderate DR, Severe DR, and Proliferative DR. In this study, we have used DenseNet121 pre-trained architecture followed by Pooling, Dropout, and SoftMax layers. Our network trained on APTOS2019 Kaggle's dataset, outperformed other state-of-the-art networks in early-stage detection, and achieved 96.51% accuracy in severity grading of Diabetic Retinopathy detection for multi-label classification and achieved 94.44% accuracy of simple single-class classification. Moreover, we have evaluated the performance of our network in terms of precision, recall, f1-score, weighted kappa, and confusion matrix.

## 3. METHODS

### 3.1 Dataset

In this work, we have used APTOS 2019 dataset [40], the most recent publicly available Kaggle dataset from the APTOS Blindness Detection competition on Kaggle for Diabetic Retinopathy Detection. The APTOS 2019 dataset contains 3662 fundus photography images which are categorized into five grades (0, 1, 2, 3, 4) according to the severity of DR which are named sequentially as Normal, Mild, moderate non-proliferative DR, severe non-proliferative DR, and proliferative DR. The sample images of Diabetic Retinopathy types from APTOS Dataset are represented in Figure 2. The dataset contains 1805 No DR images, 370 Mild DR images, 999 Moderate images, 193 Severe images, 295 Proliferative images with different resolutions, as shown in Table 1. We split the dataset of 3662 images with 3112 images in the training set and 550 images in the validation set.

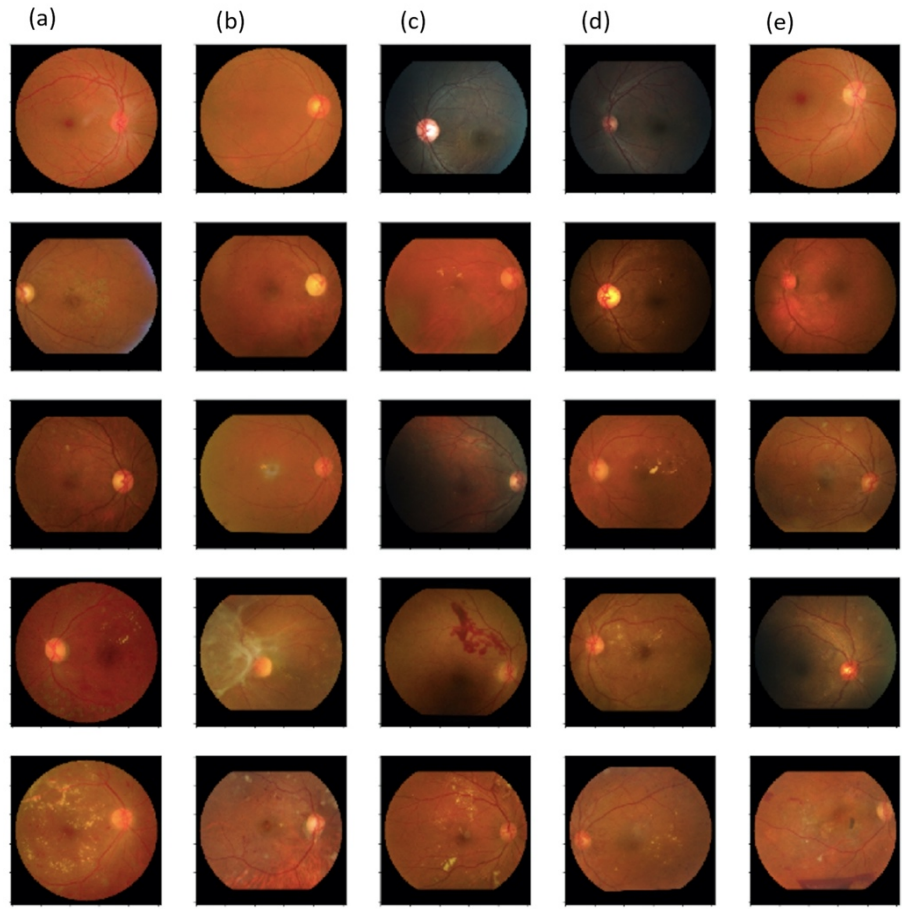

**Figure 2.** The sample fundus photography images from APTOS Dataset for Diabetic Retinopathy severity levels : (a) No DR, (b) Mild DR, (c) Moderate DR, (d) Severe DR, (e) Proliferative DR

| DR Grade | Grade Name | Total Images |
|---|---|---|
| 0 | No DR | 1805 |
| 1 | Mild DR | 370 |
| 2 | Moderate DR | 999 |
| 3 | Severe DR | 193 |
| 4 | Proliferative DR | 295 |

**Table 1.** Grade distribution of the Diabetic Retinopathy images from APTOS 2019 dataset.

### 3.2 Data Preprocessing

The preprocessing steps in the study were kept minimal to maintain better generalization of the image conditions. We performed a basic preprocessing step using the built-in preprocessing function of Keras, ImageDataGenerator [41]. The ImageDataGenerator was set to perform a random zoom of 0.15, horizontal, and vertical flip for all the images in the dataset. The fundus photography images of the dataset have a random resolution. We down-scalded the images to 224 × 224 pixels resolution to match with the DenseNet121 input image dimension.

### 3.3 Network Architecture

The Dense Convolutional Network is a network architecture where each layer is directly connected to every other layer, in a feed-forward fashion. For each layer, the feature maps of all other layers are treated as separate inputs, whereas its feature maps are passed on as inputs to all subsequent layers [42]. The advantages of DenseNets to alleviate the vanishing gradient problem, strengthen feature propagation, encourage feature reuse, and substantially reduce the number of parameters, have the edge over other Deep Convolutional Networks. DenseNets obtain significant improvements over the state-of-the-art on most of them while requiring less memory and computation to achieve high performance [43].

Table 2. illustrates the network architecture of our proposed severity grading Diabetic Retinopathy detection method. The input layer of the network is 224 x 224 pixels resolution. We have extended the architecture of DenseNet121 by the addition of GlobalAveragePooling2D having 1024 neurons, Dropout of 0.5 to reduce overfitting, and SoftMax layer having five classes.

Sigmoid [44] was used in all convolutional layers as the activation function for nonlinearity. All the Max-Pooling layers used have the same kernel size of 3 x 3. The final extracted local features were flattened before passing through fully connected layers. This architecture is fine-tuned on 3112 sample images for 15 epochs with a learning rate of 0.00005 and Adam optimizer for the faster optimization of the network.

| Layer Type | Kernel Size | Stride | Output shape |
|---|---|---|---|
| Input | - | - | (224,224,3) |
| Convolution | 7 x 7 conv | 2 | (224,224,32) |
| Pooling | 3 x 3 max pool | 2 | (112,112,32) |
| Dense Block 1 | 1 x 1 conv<br>3 x 3 conv | 6 | (56,56,64) |
| Transition Layer 1 | 1 x 1 conv<br>2 x 2 average pool | 1<br>2 | (56,56,128)<br>(28,28,128) |
| Dense Block 2 | 1 x 1 conv<br>3 x 3 conv | 12 | (28,28,256) |
| Transition Layer 2 | 1 x 1 conv<br>2 x 2 average pool | 1<br>2 | (28,28,512)<br>(14,14,512) |
| Dense Block 3 | 1 x 1 conv<br>3 x 3 conv | 24 | (7,7,1024) |
| GlobalAveragePooling | 1024 | - | (1024) |
| Dropout | 1024 | - | (1024) |
| Dense layer | 5 | - | (5) |

**Table 2.** The proposed network architecture of the severity grading Diabetic Retinopathy detection model.

### 3.4. Training

We have performed the training of our proposed network with the single-label classification method and multi-label classification method for the grading of Diabetic Retinopathy among five severity levels. In the multi-label classification, the prediction is not made on a single label; the target is set to a multi-label problem; i.e., if the target is a class 4, then it encompasses all the classes before it. The hyperparameters settings were kept constant for both single-label and multi-label classification method. Our proposed network has more than 7 million total parameters, which were randomly initialized. The network was trained with Adam optimization function [45] for a fixed schedule over 15 epochs. We tested several learning rates but found 0.00005 to be the best learning rate for the study. The summary of the settings of our training hyperparameters is given in Table 3.

| Hyperparameters | Values |
|---|---|
| Loss Function | Binary Loss |
| Optimizer | Adam |
| Learning Rate | 0.00005 |
| Batch Size | 32 |
| Alpha | 0.2 |
| Epoch | 15 |

**Table 3.** The summary of the settings of our training hyperparameters.

## 4. EXPERIMENTAL RESULTS

The calculations were performed on Kaggle kernel having 4 CPU cores with 17 GB RAM and 2 CPU cores with 14GB RAM [33]. The model evaluation was performed for both the single-label method and multi-label method used in this study by evaluating the accuracy, precision, recall, f1-score, quadratic weighted kappa indices, and confusion matrix. Also, loss and accuracy curves were plotted to keep track of the performance of the model concerning the number of epochs.

### 4.1 Performance Evaluation on Severity Grading

The validation set, which contains 550 fundus photography images, was used for model evaluation. The model was evaluated on the macro average, the weighted average for precision, recall, f1-score to know the performance of the model using the single-label classification method of the study. The macro average and the weighted average for precision, recall, and f1-score were evaluated for five classes. The macro average of 0.75, 0.67, 0.70, and the weighted average of 0.86, 0.87, 0.86 were recorded for precision, recall, and f1-score, respectively. Our model shows the best precision, recall, and f1-score value for No DR, i.e., 0.97, 0.98, 0.98, respectively. The Diabetic Retinopathy Classification system evaluation of Macro Average and Weighted Average for precision, recall, and f1-score are represented in Table 4.

| Classes | Precision | Recall | F1-support |
|---|---|---|---|
| No DR | 0.97 | 0.98 | 0.98 |
| Mild DR | 0.71 | 0.48 | 0.57 |
| Moderate DR | 0.74 | 0.95 | 0.83 |
| Severe DR | 0.75 | 0.25 | 0.38 |
| Proliferative DR | 0.82 | 0.56 | 0.67 |
| **Macro Average** | **0.75** | **0.67** | **0.70** |
| **Weighted Average** | **0.86** | **0.87** | **0.86** |

**Table 4.** Diabetic Retinopathy Classification system evaluation of Macro Average and Weighted Average for precision, recall, and f1-Score evaluated over 550 sample images of validation data using single-label classification method.

The quadratic weighted kappa, the recommended state-of-the-art performance metric for multi-class Diabetic Retinopathy, is also adopted as the performance metric of our severity grading prediction. The quadratic weighted kappa allows disagreements to be weighted differently and is especially useful when codes are ordered. The Quadratic Weighted Kappa indices for a multi-label classification method was evaluated over validation data and reported to be 91.96%. The curve of Quadratic Weighted Kappa with the number of epochs is visualized in Figure 3.

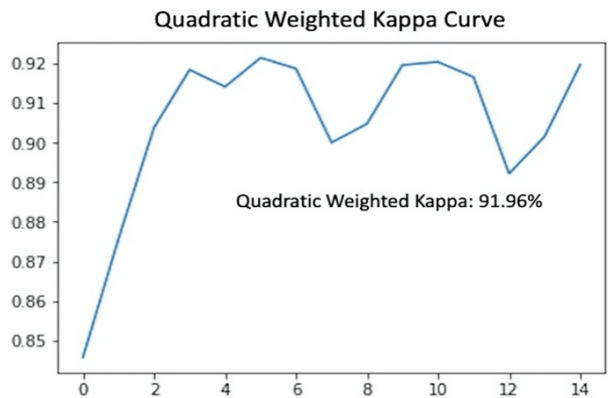

**Figure 3.** The Quadratic Weighted Kappa curve with respect to the number of epochs for a multi-label classification method. The Quadratic Weighted Kappa was determined as 91.96% evaluated on the 550 sample images of the validation data.

## 4.2 Confusion Matrix

A confusion matrix describes the performance of a classification model over a validation set by comparing the true labels with the predicted labels. The confusion matrix of our model was evaluated for the single-label classification method for five classes, which is shown in Table 5. Each element of the confusion matrix shows the comparison between the true label and predicted label for each image in the validation set. Our model showed the best result for No DR by making a correct prediction for 281 images out of 286 images. Whereas, the correct predicted images for Mild DR, Moderate DR, Severe DR, and Proliferative DR were 24, 141, 6, and 23 images out of 50, 129, 24, and 41 images, respectively. The detailed Confusion Matrix results are explained in Table 5.

| True Label / Predicted Label | No DR | Mild DR | Moderate DR | Severe DR | Proliferative DR |
|---|---|---|---|---|---|
| No DR | 281 | 4 | 1 | 0 | 0 |
| Mild DR | 6 | 24 | 19 | 0 | 1 |
| Moderate DR | 2 | 3 | 141 | 2 | 1 |
| Severe DR | 0 | 0 | 15 | 6 | 3 |
| Proliferative DR | 0 | 3 | 15 | 0 | 23 |

**Table 5.** Confusion Matrix: The diagonal elements which represent the number of points for which predicted label matches true label, while non-diagonal elements are those that are wrongly classified by the classifier.

## 4.3 Loss and Accuracy Analysis

We computed the loss and accuracy curves to keep track of the performance of our model for both the single-label and multi-label methods used in the study. Figure 4 shows the training, validation loss of the model for the single-label method, and multi-label method. The training loss and validation loss for the single-label method were found to be 0.1008 and 0.1435, respectively. Whereas the training loss and validation loss for the multi-label method were found to be 0.0682 and 0.1061, respectively. Figure 5 shows the training, validation accuracy of the model for the single-label method and multi-label method of diabetes Retinopathy classification. The training accuracy and validation accuracy for the single-label method was found to be 95.98% and 94.44%, respectively. Whereas the training accuracy and validation accuracy for the multi-label method was found to be 97.54% and 96.40%, respectively. In our study, It is evident that the training and validation curve was best reported for the multi-label method as compared to the single-label method.

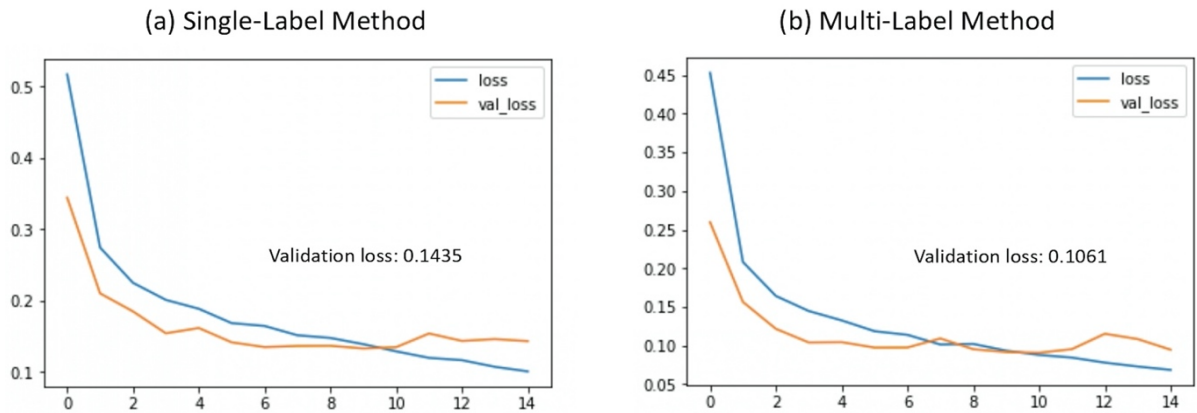

**Figure 4.** The Training Loss vs Validation Loss Curves was evaluated over the training set and validation set (a) Single-Label Method (b) Multi-Label Method

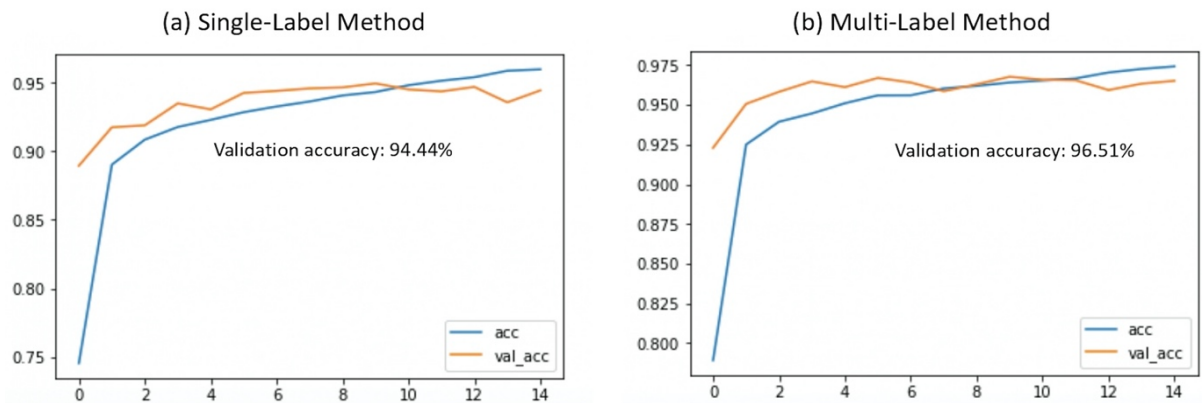

**Figure 5.** The Training Accuracy vs Validation Accuracy Curves was evaluated over the training set and validation set (a) Single-Label Method (b) Multi-Label Method

The comparison of the current study with other related previous work with the number of classes four and five were included in the study, which is represented in Table 6. The accuracies and recall for the previously reported studies vary between approximately 37 percent to 94 percent and 30 percent to 90 percent, respectively. Our methods for multi-class Diabetic Retinopathy Classification has performed better than previously proposed computer-aided diagnosis systems in terms of both accuracy and recall. We have achieved 96.51% accuracy in severity grading of Diabetic Retinopathy detection for multi-label classification and achieved 94.44% accuracy of simple single-class classification.

| Author | Year | Classifier | No. of Classes | Accuracy | Recall |
|---|---|---|---|---|---|
| [46] | 2008 | NN | 4 | 84% | 90% |
| [26] | 2011 | SVM | 5 | 82% | 82.50% |
| [34] | 2016 | CNN | 5 | 75% | 30% |
| [47] | 2016 | CNN | 5 | 94% | 76.6% |
| [48] | 2016 | GoogleNet | 5 | 45% | - |
| [49] | 2018 | GoogleNet and AlexNet | 4 | 57.2% | - |
| [50] | 2018 | AlexNet | 5 | 37.43% | - |
| | | VGG16 | | 50.03% | - |
| | | InceptionNet V3 | | 63.23% | - |
| [35] | 2018 | Ensemble of CNN | 4 | 83.9% | - |
| Current Study | 2020 | DenseNet121 (Single label) | 5 | **94.44%** | 87% |
| | | DenseNet121 (Multi label) | | **96.51%** | - |

**Table 6.** The comparison of the current study with other related previous work with number of classes four and five. The current study achieved 96.51% accuracy in severity grading of Diabetic Retinopathy detection for multi-label classification and achieved 94.44% accuracy, recall 87% for simple single-label classification method.

## 5. CONCLUSION

In this paper, we have presented a DenseNet121 model for early-stage detection of five severity grades for Diabetic Retinopathy. We performed several modifications in the pre-trained DenseNet121 network and employed preprocessing to improve the performance of the network. Our network was trained on APTOS 2019 dataset, which outperformed other state-of-the-art networks in early-stage detection. The best accuracy of 96.51% was reported with the multi-label classification method in comparison of 94.44% accuracy of the single-label classification method. Also, the precision, recall, f1-score, and quadratic weighted kappa for our network was reported as 86%, 87%, 86%, and 91.96%, respectively. In our work, we have found that without heavy data preprocessing and data augmentation, well-tuned hyperparameters for the network can be effectively used to perform training on training data. Even without data-augmentation, the proposed method has achieved a better performance than previously reported computer-aided Diabetic Retinopathy detection systems. The experimental results have demonstrated the effectiveness of our proposed method to be good enough to be employed in clinical applications for Diabetic Retinopathy Detection.